\documentclass[sigconf,screen]{acmart}
\AtBeginDocument{%
  \providecommand\BibTeX{{%
    \normalfont B\kern-0.5em{\scshape i\kern-0.25em b}\kern-0.8em\TeX}}}

\setcopyright{acmcopyright}
\copyrightyear{2022}
\acmYear{2022}
\acmDOI{10.1145/3503161.3548397}

\def\Vec#1{\textit{\boldmath $#1$}}

\usepackage{balance}
\usepackage{hyperref}           
\hypersetup{
    colorlinks=true,
    linkcolor=blue,
    filecolor=red,      
    urlcolor=magenta,
    breaklinks=true,            
}
\usepackage{breakurl}           

\acmConference[MM '22] {Proceedings of the 30th ACM International Conference on Multimedia}{October 10--14, 2022}{Lisboa, Portugal}
\acmBooktitle{Proceedings of the 30th ACM International Conference on Multimedia (MM '22), October 10--14, 2022, Lisboa, Portugal}

\acmPrice{15.00}
\acmISBN{978-1-4503-9203-7/22/10}

\begin{document}

\title{ConceptBeam: Concept Driven Target Speech Extraction}

\author{Yasunori Ohishi}
\affiliation{%
  \institution{NTT Corporation, Tokyo, Japan}
  \city{}
  \country{}
}
\email{yasunori.ooishi.uk@hco.ntt.co.jp}
\author{Marc Delcroix}
\affiliation{%
  \institution{NTT Corporation, Tokyo, Japan}
  \city{}
  \country{}
}
\email{marc.delcroix.hc@hco.ntt.co.jp}
\author{Tsubasa Ochiai}
\affiliation{%
  \institution{NTT Corporation, Tokyo, Japan}
  \city{}
  \country{}
}
\email{tsubasa.ochiai.ah@hco.ntt.co.jp}
\author{Shoko Araki}
\affiliation{%
  \institution{NTT Corporation, Tokyo, Japan}
  \city{}
  \country{}
}
\email{shoko.araki.pu@hco.ntt.co.jp}
\author{Daiki Takeuchi}
\affiliation{%
  \institution{NTT Corporation, Tokyo, Japan}
  \city{}
  \country{}
}
\email{daiki.takeuchi.ux@hco.ntt.co.jp}
\author{Daisuke Niizumi}
\affiliation{%
  \institution{NTT Corporation, Tokyo, Japan}
  \city{}
  \country{}
}
\email{daisuke.niizumi.dt@hco.ntt.co.jp}
\author{Akisato Kimura}
\affiliation{%
  \institution{NTT Corporation, Tokyo, Japan}
  \city{}
  \country{}
}
\email{akisato.kimura.xn@hco.ntt.co.jp}
\author{Noboru Harada}
\affiliation{%
  \institution{NTT Corporation, Tokyo, Japan}
  \city{}
  \country{}
}
\email{noboru.harada.pv@hco.ntt.co.jp}
\author{Kunio Kashino}
\affiliation{%
  \institution{NTT Corporation, Tokyo, Japan}
  \city{}
  \country{}
}
\email{kunio.kashino.me@hco.ntt.co.jp}

\renewcommand{\shortauthors}{Yasunori Ohishi et al.}

\begin{abstract}
We propose a novel framework for target speech extraction based on semantic information, called \textit{ConceptBeam}.
Target speech extraction means extracting the speech of a target speaker in a mixture. Typical approaches have been exploiting properties of audio signals, such as harmonic structure and direction of arrival. 
In contrast, ConceptBeam tackles the problem with semantic clues. Specifically, we extract the speech of speakers speaking about a concept, i.e., a topic of interest, using a concept specifier such as an image or speech.
Solving this novel problem would open the door to innovative applications such as listening systems that focus on a particular topic discussed in a conversation.
Unlike keywords, concepts are abstract notions, making it challenging to directly represent a target concept.
In our scheme, a concept is encoded as a semantic embedding by mapping the concept specifier to a shared embedding space.
This modality-independent space can be built by means of deep metric learning using paired data consisting of images and their spoken captions.
We use it to bridge modality-dependent information, i.e., the speech segments in the mixture, and the specified, modality-independent concept.
As a proof of our scheme, we performed experiments using a set of images associated with spoken captions. That is, we generated speech mixtures from these spoken captions and used the images or speech signals as the concept specifiers.
We then extracted the target speech using the acoustic characteristics of the identified segments.
We compare ConceptBeam with two methods: one based on keywords obtained from recognition systems and another based on sound source separation.
We show that ConceptBeam clearly outperforms the baseline methods and effectively extracts speech based on the semantic representation.
\end{abstract}


\maketitle
\begin{figure}[h]
  \vspace{-4mm}
  \centering
  \includegraphics[width=\linewidth]{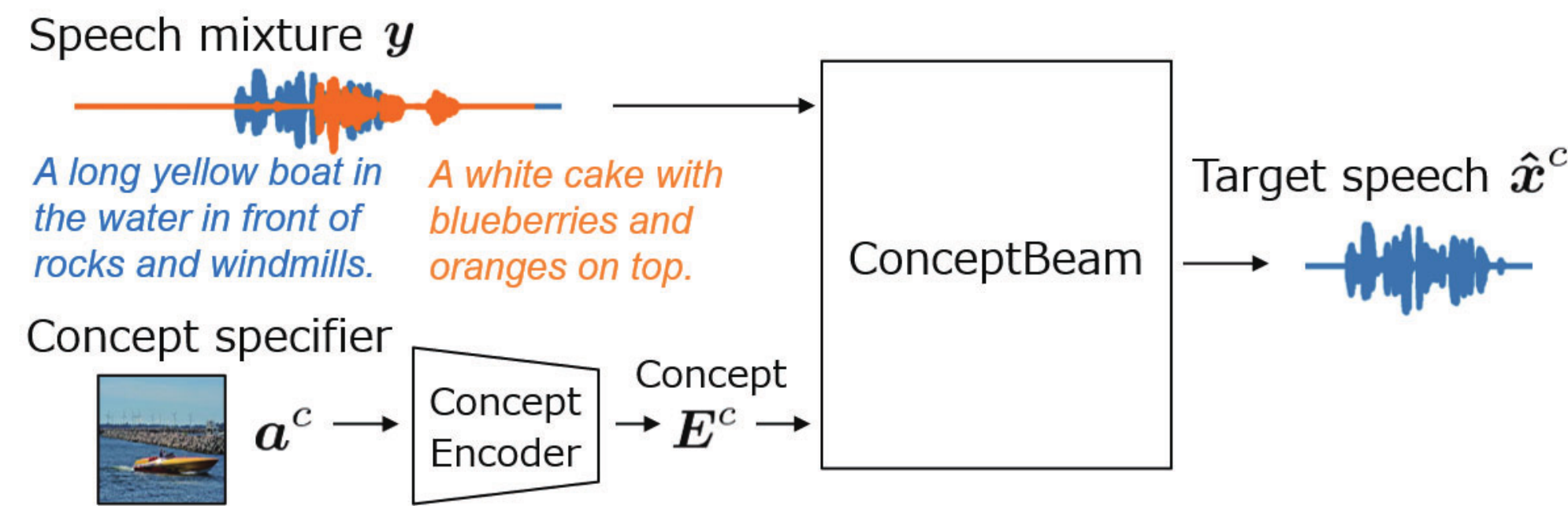}
  \caption{Concept driven target speech extraction.}
  \label{fig:figure1}
  \vspace{-5mm}
\end{figure}
\section{Introduction}
\label{sec:intro}


Target speech extraction is one approach towards achieving the cocktail party effect or selective hearing, which refers to our ability to focus on a specific auditory stimulus while filtering out others.
Research on target speech extraction has conventionally focused on extracting the speech of a target speaker in a mixture of overlapping speakers by exploiting physical clues such as pre-recorded enrollment utterances \cite{Zmolikova2019}, direction information \cite{Gu2019}, or video \cite{Afouras2018,Ephrat2018,Ochiai2019,Sato2021} to identify the \emph{target speaker}.
Meanwhile, we can use \emph{semantic clues}, such as language \cite{9688052,tzinis2022heterogeneous} or content of speech, to focus our attention on the conversation we want to hear.
For example, if our name is mentioned or the topic of a conversation nearby interests us, we turn our attention to that speaker.

We propose a novel framework called \textit{ConceptBeam} for target speech extraction based on ``concept'' or semantic information.
In the context of this paper, a concept is an abstract term that refers to the semantic content of an utterance. For instance, it could be the topic or theme mentioned in the utterance and is a broader notion than keywords.
ConceptBeam uses a concept specifier such as an image or speech to tackle the problem of extracting the speech of speakers speaking about a concept.
Solving this novel problem would open the door to innovative listening systems that focus on a particular topic discussed in a conversation.

However, since a concept is an abstract notion, it is challenging to define and directly represent a target concept.
In our proposed scheme, a concept is encoded as a semantic embedding by mapping the concept specifier to a shared embedding space \cite{Merkx2019,Harwath2019b,Harwath2020,Ohishi2020a,Miech2020,peng2022fastvgs}.
This modality-independent space can be built by means of deep metric learning using paired data consisting of images and their spoken captions.
We use it to bridge modality-dependent information, that is, the speech segments in the mixture, and the specified, modality-independent concept.
Once we obtain the semantic representations of concepts, we can identify the speech segments in the mixture related to them. Finally, we extract the target speech using the acoustic characteristics estimated from the identified segments in a manner similar to the activity driven extraction network (ADEnet) \cite{Delcroix2021}, which exploits speaker activity information for target speech extraction.

Figure~\ref{fig:figure1} illustrates ConceptBeam when an image is used as a concept specifier.
The input speech mixture consists of two utterances about different concepts (i.e., one describing a boat and one a cake).
Suppose we are interested in listening to what is said about the boat. We can then use an image of a boat as a concept specifier to define the target concept.
We derive a semantic embedding from the concept specifier using a concept encoder, as we described above.
ConceptBeam involves the use of neural networks (NNs) and accepts the mixture and semantic embedding and outputs only the speech related to the target concept.



We conducted experiments in which we regarded spoken descriptions of images (i.e., spoken captions) as examples of speech about concepts. We generated speech mixtures from these spoken captions and used the images or speech signals as the concept specifiers. 
We then compared the ConceptBeam with two baseline methods: one based on keywords obtained from object and speech recognition systems and another based on sound source separation.
These methods are susceptible to modality-specific recognition errors and expression variations in the recognition systems, as well as to interfering speech, making speech extraction difficult.
We show that ConceptBeam can effectively extract speech based on the modality-independent semantic embedding, thus clearly outperforming the baseline methods.


\vspace{-2mm}
\section{Related work}
\label{sec:related_work}

The problem with target speech extraction is related to speech separation \cite{Hershey2016,Kolbak2017}.
For example, as an alternative to ConceptBeam, we could develop a similar semantic-based extraction system by first separating speech signals into all speech sources and then identifying the target speech among the separated signals using a concept specifier. In fact, we use such a separation-based approach for comparison in our experiments. 
Target speech extraction has been proposed as an alternative to speech separation to extract directly the target speech signal from a mixture.
This requires a clue to inform the system about the target \cite{Zmolikova2017asru,Zmolikova2019,chen2018deep,Gu2019,Afouras2018,Ephrat2018,Ochiai2019,Sato2021,Delcroix2021,Ceolini2020}.

Our work is also related to visual sound separation \cite{Zhao2018,Xu2019,Gao2019,Zhao2019,Gan2020}.
These works demonstrated how semantic information contained in appearances and motions could help in sound separation.
Unlike these works, we directly use semantic context or concept clues obtained from audio-visual learning.

Many researchers have proposed unsupervised spoken language learning methods that can learn meaningful audio-visual correspondences from visually grounded speech \cite{Harwath2016,Harwath2020,Chrupala2021,Monfort2021,Rouditchenko2021b,Palmer2021,Wang2021,scholten2021learning,Sanabria2021,olaleye2021attention,Peng2022word,peng2022fastvgsplus,mitja2022learning}.
\cite{Harwath2016} proposed a multiple-encoder model to associate visual objects with spoken words and showed a similarity profile between an image and its spoken caption.
The similarity profile is based on the dot product of the image embedding vector with the audio embedding vector obtained from the multiple-encoder model.
Since the regions of the waveform with the highest similarity to the image corresponds to highly informative words or phrases that explicitly refer to the salient objects, the semantic embeddings may represent a type of semantic concept.
ConceptBeam was inspired by this work.
Unlike \cite{Harwath2016}, which used embeddings for a crossmodal retrieval task, we exploit these semantic embeddings to extract speech, though other means to represent concepts such as keywords and topics could also be used.

\begin{figure}[t]
  \centering
  \includegraphics[width=\linewidth]{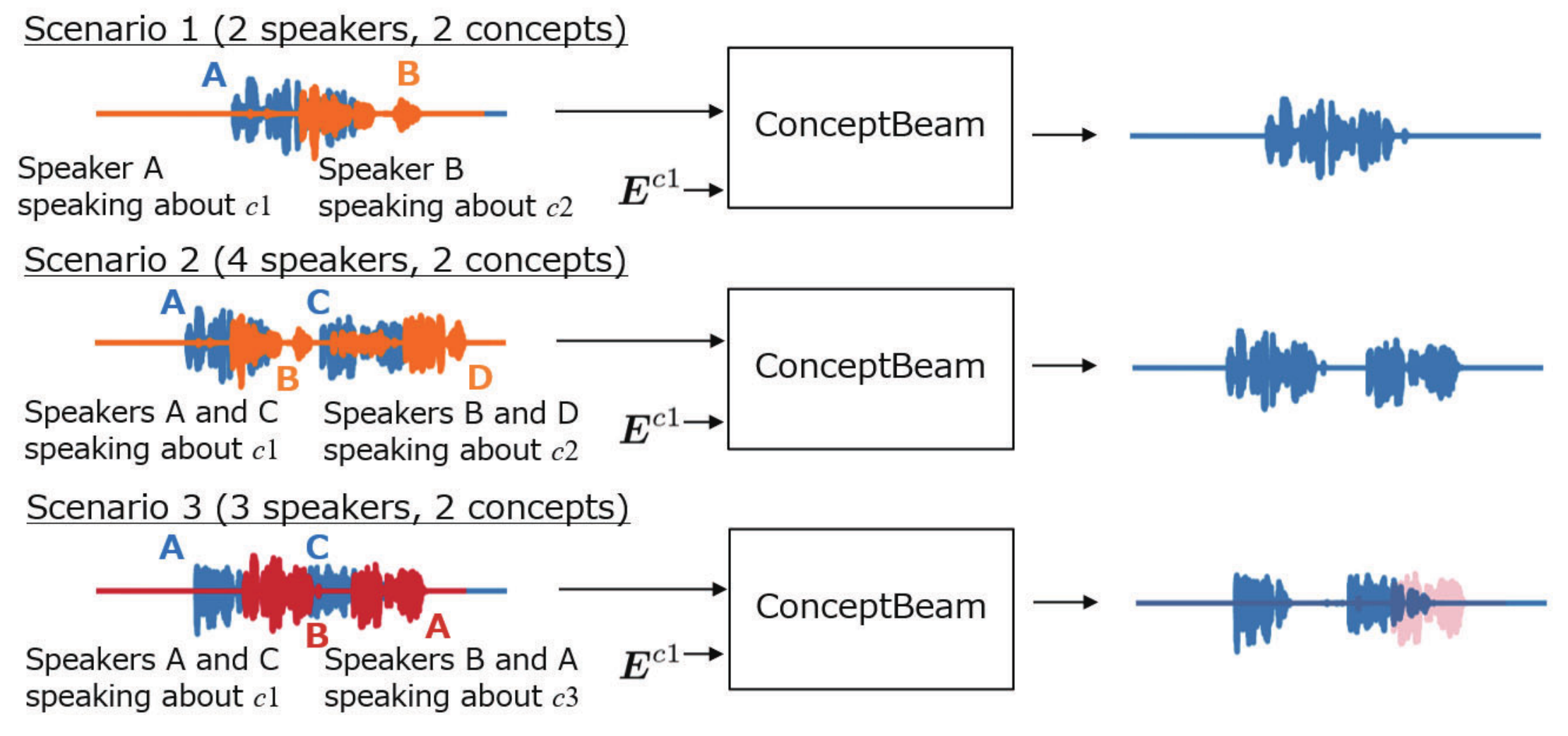}
  \vspace{-6mm}
  \caption{Different scenarios of concept driven target speech extraction.
  }
  \label{fig:cases}
  \vspace{-6mm}
\end{figure}
\section{Concept driven target speech extraction}
\label{sec:problem_formulation}

\subsection{Problem formulation}

We consider the problem of extracting the speech of one or more speakers speaking about a particular concept from a speech mixture captured by a single microphone.
The observed signal is given as
\begin{align}
\Vec{y}=\sum_{m=1}^C\sum_{n=1}^{N}\Vec{x}^{(m,n)}+\Vec{b},
\end{align}
where $\Vec{y}\in\mathbb{R}^T$, $\Vec{x}^{(m,n)}\in\mathbb{R}^T$, and $\Vec{b}\in\mathbb{R}^T$ are the observed single-channel mixture, source signal from the $n$th speaker speaking about the $m$th concept, and background noise, respectively,
$N$ and $C$ are the number of speakers and concepts in the mixture, respectively, and $T$ is the signal duration.
We consider that $\Vec{x}^{(m,n)}\!=\!\Vec{0}$ when the $n$th speaker is not speaking about the $m$th concept, where $\Vec{0}$ denotes a vector of all zeros.

The goal of concept driven target speech extraction is to extract the target signal, $x^{c}$, which consists of the sum of the speech signals of the speakers speaking about the target concept $c$,
\begin{align}
\Vec{x}^{c}=\sum_{n=1}^N\Vec{x}^{(c,n)}.
\end{align}
Note that this requires an auxiliary input for specifying the target concept.
We represent a concept specifier that is assumed to contain the concept we want to extract as $\Vec{a}^{c}$.
In this paper, we consider two types of concept specifiers: an image and speech. For example, if we wish to extract a conversation about sweets, we could input an image of a popular sweet or an utterance describing sweets.

\subsection{Typical scenarios}
Figure~\ref{fig:cases} illustrates the three different scenarios of concept driven target speech extraction. 
In Scenario 1, two speakers (A and B) speak about two different concepts ($c1$ and $c2$).
Given a semantic embedding $\Vec{E}^{c1}$, ConceptBeam should extract the speech of speaker A speaking about $c1$.
In Scenario 2, a mixture of four different speakers (A, B, C, and D) is received, with A and C speaking about the same concept, i.e., $c1$, and B and D about a different concept, i.e., $c2$.
In this scenario, given a semantic embedding $\Vec{E}^{c1}$, ConceptBeam should extract the speech signals related to $c1$, the speech signals of speakers A and C.
Finally, in Scenario 3, speakers A and C speak about one concept, i.e., $c1$, and speakers A and B about another concept, i.e., $c3$, i.e., speaker A speaks about the two concepts.
In this case, ConceptBeam should extract the speech signals related to $c1$, i.e., the speech of speakers A and C, and filter out the speech of speaker A when A speaks about the other concept, i.e., $c3$.
As presented above, ConceptBeam should focus on the coincidence of concepts and ignore other properties such as speakers.

\subsection{Concept representation}
\label{ssec:concept_representation}
\begin{figure}[t]
  \centering
  \includegraphics[width=0.90\linewidth]{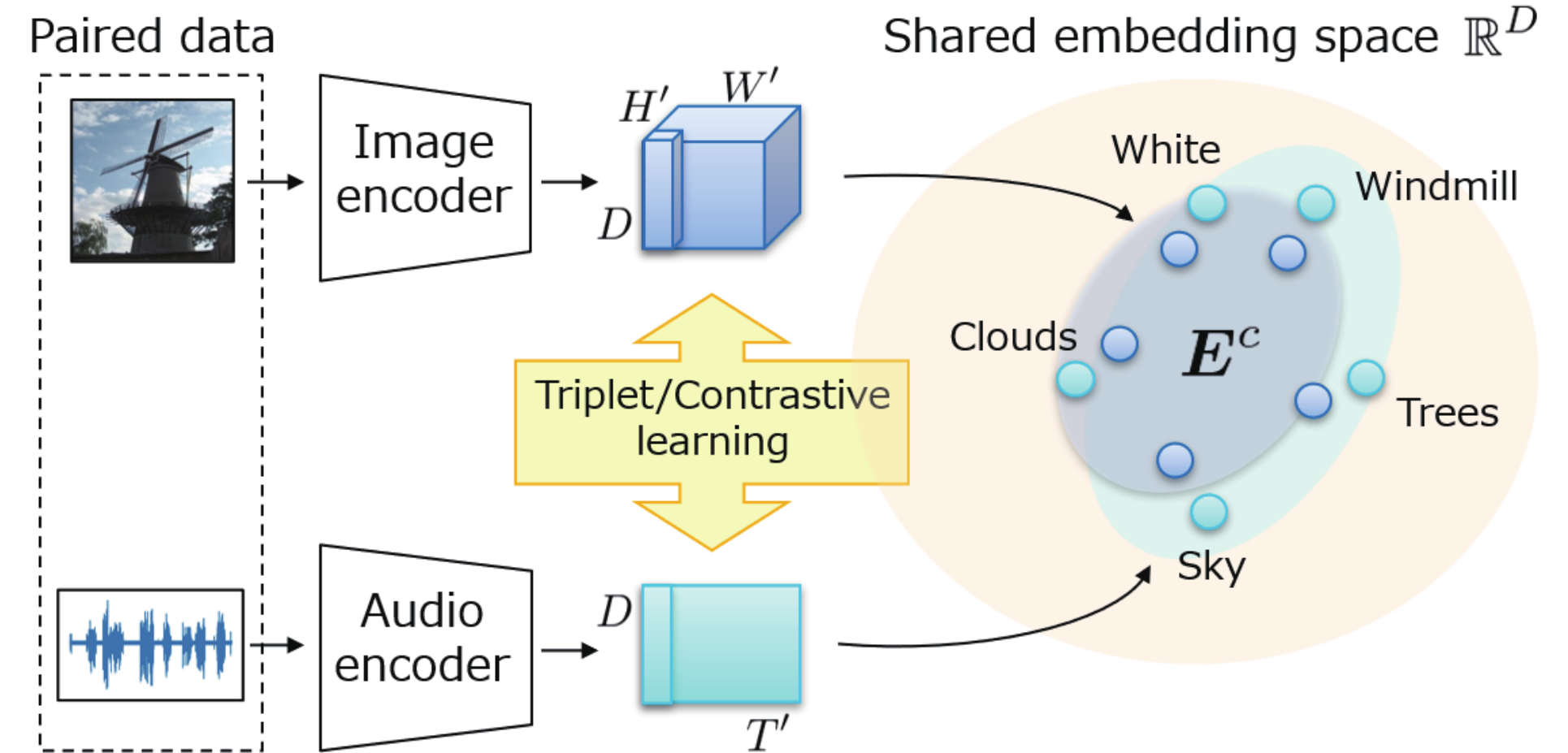}
  \caption{Modality-independent semantic representation.}
  \label{fig:figure3}
  \vspace{-3mm}
\end{figure}

We extract a semantic embedding, $\Vec{E}^{c}$, from a concept specifier using an image or audio encoder of a multiple-encoder model \cite{Ilharco2019,Harwath2019b,Sanabria2021}.
These encoders are trained with deep metric learning using paired data consisting of images and their spoken captions.
They learn a shared embedding space by mapping close to each other the embedding obtained from an image and the embedding obtained from its spoken caption.
Figure~\ref{fig:figure3} illustrates the training process for learning a modality-independent semantic representation.
A concept is defined as a closed domain formed by $\Vec{E}^{c}$ in the modality-independent embedding space.
We use a pre-trained multiple-encoder model~\cite{Harwath2019b}, with which the semantic embedding is $\Vec{E}^{c} \!\in\! \mathbb{R}^{D\times K}$, $K\!=\!H'\!*\!W'$ for an image concept specifier and $K\!=\!T'$ for a speech concept specifier.
The notations $D$, $H'$, $W'$, and $T'$ respectively denote the dimension of the embedding, the downsampled height and width of the image, and the downsampled number of time frames.

\section{ConceptBeam framework}
\begin{figure}[t]
  \centering
  \includegraphics[width=0.87\linewidth]{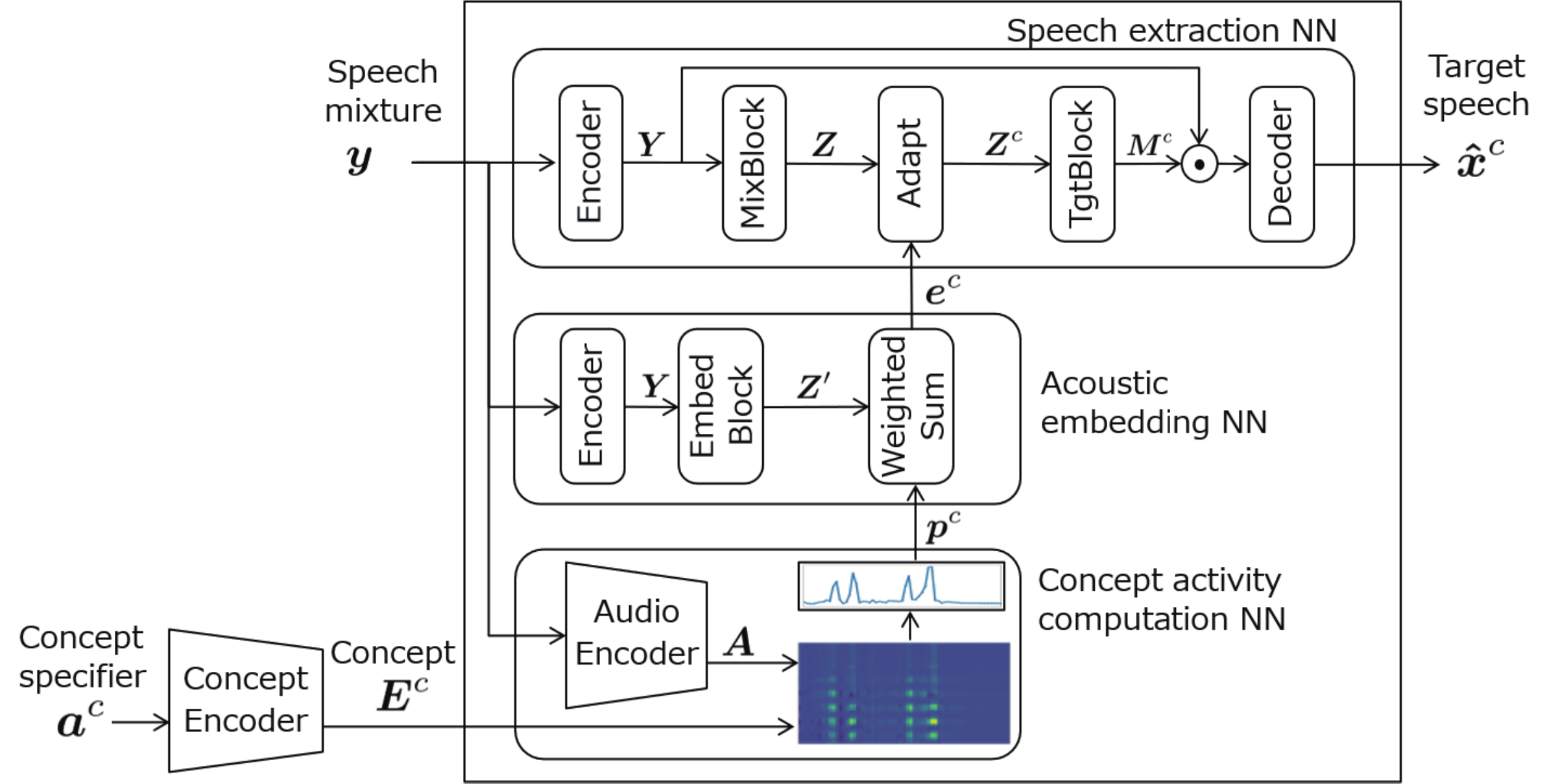}
  \caption{Diagram of ConceptBeam.}
  \label{fig:Diagram}
  \vspace{-4mm}
\end{figure}

ConceptBeam is composed of four modules: a concept encoder, concept activity computation NN, acoustic embedding NN, and speech extraction NN, as shown in Figure~\ref{fig:Diagram}.
The proposed ConceptBeam is related to the activity driven extraction network (ADEnet), which exploits speaker activity information for target speech extraction \cite{Delcroix2021}.
ConceptBeam uses a concept specifier to determine the target activity.
Unlike ADEnet, the activity may cover multiple speakers speaking about the same concept.

\subsection{Concept encoder}
The concept encoder computes the semantic embedding that represents the target concept from a concept specifier as
\begin{align}
    \Vec{E}^c = \mathrm{ConceptEncoder}(\Vec{a}^c),
\end{align}
where we use the image or audio encoder of the multiple-encoder model described in Section \ref{ssec:concept_representation} depending on whether the concept specifier is image or speech.

\subsection{Concept activity computation neural network}
This module infers a concept activity that identifies the speech segments in the input mixture related to the target concept as
\begin{align}
    \Vec{p}^c = f(\Vec{y}, \Vec{E}^c),
\end{align}
where $\Vec{p}^c\!=\{p_1^c,p_2^c,\ldots,p_{T''}^c\}\!\in\!\mathbb{R}^{T''}$ is the concept activity.
With the multiple-encoder model, we can learn to map images and speech that are semantically related to a shared embedding space.
Similarly, semantically related speech signals can be mapped close together.
We use this property to find the regions of the speech related to the concept by measuring the similarity between the embedding obtained from the speech mixture using the audio encoder of the multiple-encoder model and that from the concept specifier.
The $f(\cdot)$ is this module implemented as
\begin{align}
    \Vec{A} &= \mathrm{AudioEncoder}(\Vec{y}), \\
    \Vec{P}^c &=  {\Vec{E}^c}^{\mathrm{T}}\Vec{A},\\
    p_t^c &= \max \Vec{P}_t^c, \label{eq:concept_activity}
\end{align}
where $\Vec{A}\!\in\!\mathbb{R}^{D\times T''}$ is the embedding obtained from the input speech mixture.
$T''$ denotes the number of time frames downsampled by the audio encoder, and $\Vec{P}^c\in\mathbb{R}^{K\times T''}$ is called a ``matchmap'', which indicates semantic alignments between acoustic frames and image pixels \cite{Harwath2019b}.
Here, this matrix also infers semantic alignments between acoustic frames for the speech mixture and the concept specifier.
Finally, by taking the max operation over the row axis of the matchmap, we obtain for each time frame a value indicating how close the semantic content of that speech frame is to the concept specifier, where $\Vec{P}_t^c$ is the $t$th time frame of $\Vec{P}^c$.

\subsection{Acoustic embedding neural network}
The acoustic embedding NN computes an embedding vector that represents the acoustic characteristics of the speech segments identified by the concept activity.
A prior study \cite{Delcroix2021} revealed that the speech of a speaker can be extracted from the speaker activity by deriving a speaker embedding vector from the mixture and speaker activity.
We rely on a similar idea but extend the embedding to represent more broadly the acoustic condition of the utterances related to the concept specifier.
For example, the acoustic embedding may represent the characteristics of multiple speakers that speak about the concept.
We use a sequence summary NN \cite{Vesely2016,Zmolikova2019} to convert the speech mixture to a vector of a fixed dimension, $\Vec{e}^c\in \mathbb{R}^{D'}$, as
\begin{align}
    \Vec{e}^c = g(\Vec{y}, \Vec{p}^c),
\end{align}
where $g(\cdot)$ is the acoustic embedding NN implemented as
\begin{align}
    \Vec{Y} &= \mathrm{Encoder}(\Vec{y}),  \label{encoder}\\
    \Vec{Z}' &= \mathrm{EmbedBlock}(\Vec{Y}),  \\
    \Vec{e}^c&=\frac{1}{T'''}\sum_{t=1}^{T'''}p^c_t\Vec{Z}'_t.
\end{align}
Here, $\mathrm{Encoder(\cdot)}$ is an encoder layer consisting of a short-time Fourier transform (STFT), $\mathrm{EmbedBlock}(\cdot)$ consists of two fully connected (FC) layers and a self-attention layer, $\Vec{Y}\in\mathbb{R}^{F\times T'''}$ and $\Vec{Z'}\in\mathbb{R}^{D'\times T'''}$ are internal representations of the input speech sample, $\Vec{Z}'_t$ is the $t$th time frame of $\Vec{Z}'$, and $T'''$ and $F$ are the number of time frames and frequency bins, respectively.
The last layer performs a weighted sum over the time dimension, where $p_t^c$ is linearly interpolated to align frame indices with $\Vec{Z}'_t$.

\subsection{Speech extraction neural network}
The speech extraction NN estimates the target signal from the mixture and the acoustic embedding vector as
\begin{align}
    \Vec{\hat{x}}^c = h(\Vec{y}, \Vec{e}^c),
\end{align}
where $h(\cdot)$ is the speech extraction NN implemented as
\begin{align}
    \Vec{Y} &= \mathrm{Encoder}(\Vec{y}),\\
    \Vec{Z} &= \mathrm{MixBlock}(\Vec{Y}),\\
    \Vec{Z}^c &= \mathrm{Adapt}(\Vec{Z}, \Vec{e}^c),\\
    \Vec{M}^c &= \mathrm{TgtBlock}(\Vec{Z}^c), \\
    \Vec{\hat{x}}^c &= \mathrm{Decoder}(\Vec{M}^c\odot \Vec{Y}),
\end{align}
where $\mathrm{Encoder}(\cdot)$ is the same encoder layer as Eq. (\ref{encoder}), $\Vec{Z}\in\mathbb{R}^{D'\times T'''}$, and $\Vec{Z^c}\in\mathbb{R}^{D'\times T'''}$ are the internal representations of the mixture and target, respectively;
$\mathrm{MixBlock(\cdot)}$ is the lower part of the extraction NN, which transforms $\Vec{Y}$ into a general internal representation of the mixture independent of the target concept;
$\mathrm{TgtBlock(\cdot)}$ is the upper part of the extraction NN, which computes the target time-frequency mask, $\Vec{M}^c$, that extracts the speech speaking about the target concept; and $\mathrm{MixBlock(\cdot)}$ and $\mathrm{TgtBlock(\cdot)}$ are implemented as stacks of bidirectional long short-term memory (BLSTM) layers.
$\mathrm{Decoder(\cdot)}$ is a decoder layer consisting of an inverse STFT.

The $\mathrm{Adapt(\cdot)}$ is an adaptation layer that combines $\Vec{Z}$ with the acoustic embedding $\Vec{e}^c$.
We used the simple element-wise multiplication \cite{Zmolikova2019} as $\Vec{Z}^c = \Vec{Z}\odot \Vec{e}^c$.


\subsection{Training ConceptBeam}
We require the speech mixture $\Vec{y}$, target speech signal $\Vec{x}^{c}$, and concept specifier $\Vec{a}^{c}$ to train ConceptBeam. 
We use the negative signal-to-distortion ratio (SDR) as training loss,
\begin{align}
\mathcal{L}(\Vec{\hat{x}}^c, \Vec{x}^c)=-10\log_{10}(\frac{||\Vec{x}^c||^2}{||\Vec{x}^c-\Vec{\hat{x}}^c||^2}).
\end{align}
Note that we fix the parameters of the pre-trained audio and image encoders for computing the semantic embeddings and concept activity. We train all other network parameters from scratch.

We train different ConceptBeam models for each concept specifier (image or speech).
During training, we used the oracle concept activity obtained by feeding the reference target speech signal into the acoustic embedding NN and concept activity computation NN instead of the input signal.
We expect that the oracle concept activity allows the models to better capture the characteristics of the target concepts. Details on training in our experimentation are described in the following section.


\begin{figure}[t]
  \centering
  \includegraphics[width=\linewidth]{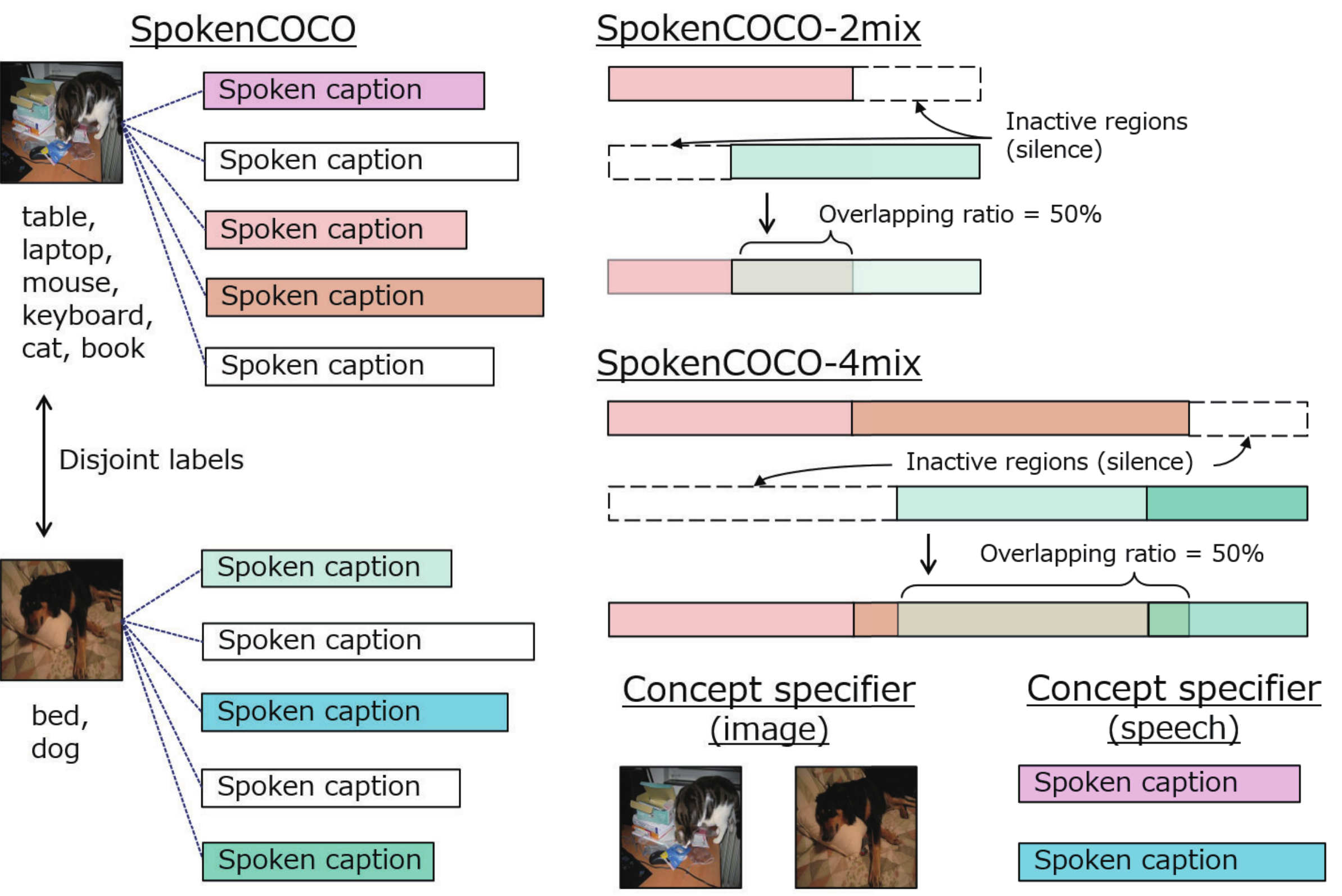}
   \caption{Preparation of SpokenCOCO-2mix and SpokenCOCO-4mix datasets.
  }
  \label{dataset}
  \vspace{-5mm}
\end{figure}

\section{Experiments}
\label{sec:experiments}


\subsection{Dataset}
SpokenCOCO \cite{Hsu2020} contains approximately 600,000 recordings of human speakers reading the MSCOCO image captions \cite{lin2014microsoft} out loud in English, where each MSCOCO caption is read once.
To represent Scenario 1, in which two speakers speak about different concepts, and Scenario 2, in which two speakers speak about the same concept and two interfering speakers speak about another concept, as shown in Figure~\ref{fig:cases}, we created two datasets, SpokenCOCO-2mix and SpokenCOCO-4mix for Scenario 1 and Scenario 2, respectively, based on spoken captions taken from SpokenCOCO.

\begin{figure*}[t]
  \centering
  \includegraphics[width=0.9\linewidth]{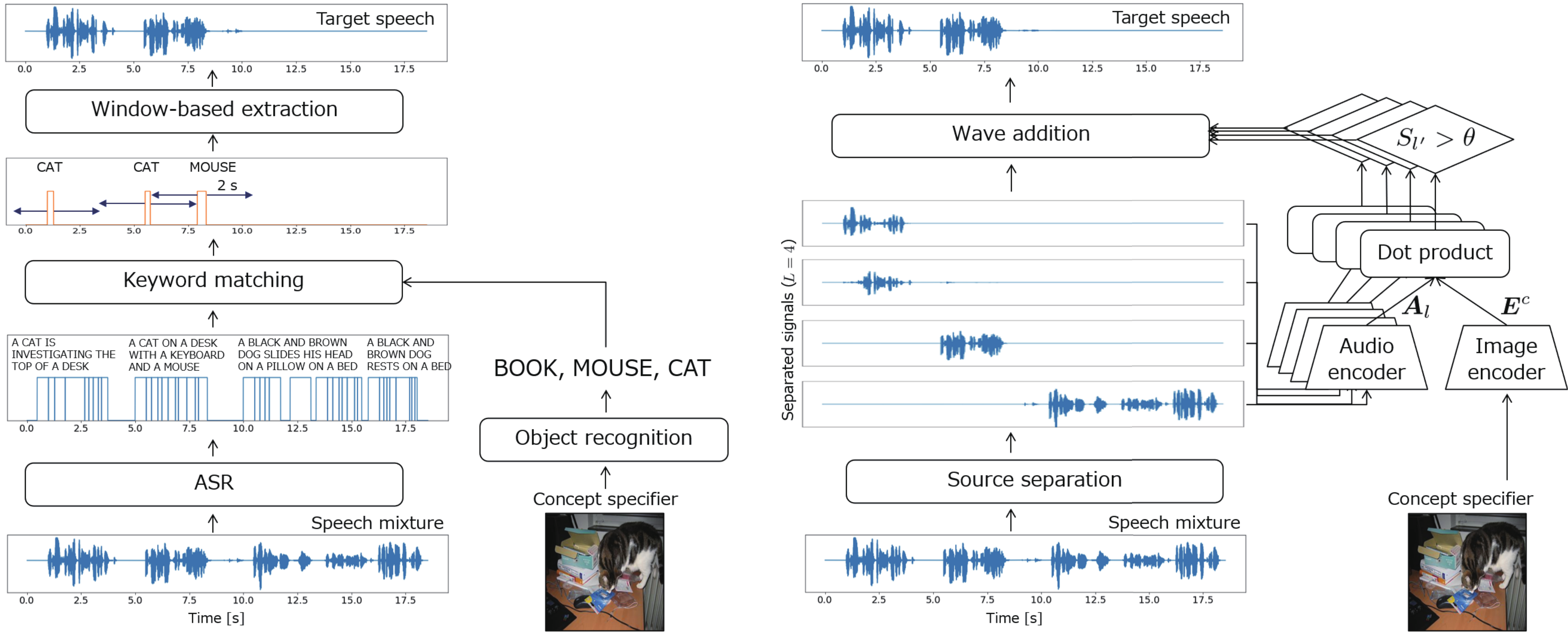}
  \caption{Keyword-based method (left) that extracts speech based on keywords obtained from modality-specific object and speech recognition systems. Separation-based method  (right) that separates the speech mixture into all of its source signals and determines speech signals containing the target concept among separated signals.}
  \label{Baseline_and_separation_based_method}
  \vspace{-3mm}
\end{figure*}

As shown in Figure ~\ref{dataset}, each image is labeled with object categories and has five spoken captions describing the content.
We mixed the signals at different overlapping ratios, the duration of the overlapping region divided by the duration of the target speech.
SpokenCOCO-2mix contains mixtures of two utterances, where we randomly selected two images with disjoint object labels, randomly chose one spoken caption for each image, and mixed the selected spoken captions at a signal-to-interference ratio between 0 and 5 dB.
SpokenCOCO-4mix contains mixtures of four utterances, where we randomly selected two spoken captions for each of the two images, concatenated the spoken captions of each image, and then mixed the signals.
Note that the generated speech mixture may contain captions spoken by the same speaker.
In the training and validation sets, the signals were mixed at overlapping ratio of 100\%, and in the testing set, they were mixed at four different overlapping ratios of 0, 25, 50, and 100\%.
The image concept specifier is the image paired with the spoken caption used to generate the mixture.
The speech concept specifier is another spoken caption paired with the image.

The training set contains 99,000 mixtures from 2,052 speakers, the validation set contains 2,137 mixtures from the same 2,052 speakers, and the evaluation set contains 3,322 utterances from 300 speakers (unseen in the training).
We downsampled the spoken captions to 8 kHz to reduce computational and memory costs.

\subsection{Network configuration}
As the audio features, we used 258-dimensional vectors of concatenated real and imaginary parts obtained using an STFT with a window of 32 ms and a shift of 8 ms.
For the image features, we pre-processed the images by resizing the smallest dimension to 256 pixels, taking a 224~$\times$~224 center crop, and normalizing the pixels according to a global pixel mean and variance.

We used ResNet50 and ResDavenet \cite{Harwath2019b} as the image and audio encoders, respectively.
ResNet50 outputs a $7\!\times\!7$ embedding across 1,024 channels ($H'\!=\!W'\!=\!7$) for an image.
ResDAVEnet outputs a $T'$~$\times$~1,024 embedding ($D\!=\!1,024$) for the 40 log Mel filter bank spectrograms computed from input audio features.
The final temporal resolution $T'$ is 1/16 of $T$.
We pretrained these encoders with the same configuration as in the study by Harwath et al. \cite{Harwath2019b}.
To evaluate the correctness of the shared embedding space, we used the Recall at 10 scores for an image and audio caption retrieval task: image-to-audio retrieval (0.785) and audio-to-image retrieval (0.758).
We believe that our shared embedding space is correctly learned because the scores are above the chance rate of 0.01 and there is no bias in the bi-directional retrieval results (retrieval using image or audio achieves a similar level of performance).
Recently, Transformer-based encoder models with the Recall at 10 scores exceeding 0.9 have been proposed \cite{peng2022fastvgs,Sanabria2021}.
It would be worthwhile to introduce such a more correct embedding space into ConceptBeam.

The EmbedBlock consists of two FC layers with 200 and 896 hidden units, rectified linear unit (ReLU) activations, and a self-attention layer.
Thus, the acoustic embedding vector has 896 dimensions ($D'=896$).
The MixBlock and TgtBlock consist of one and three BLSTM layers with 896 units, respectively.
Each BLSTM layer was followed by a linear projection layer with 896 units to combine the forward and backward LSTM outputs.
We employed one FC layer to output a spectral mask estimated by a ReLU activation.
We used Adam \cite{Kingma2015} with an initial learning rate of 0.0001 and gradient clipping \cite{Pascanu2013}.
We stopped the training procedure after 200 epochs.

\subsection{Evaluation metrics}
We measured the extraction performance in terms of scale-invariant SDR computed with the BSSEval toolkit \cite{Vincent2006}, and report the SDR improvement (SDRi) compared with the mixtures.
SDR is defined in decibels as $10* \log_{10}(P_s/P_d)$, where $P_s$ is the power of reference (clean) target speech and $P_d$ is the power of distortions in the separated signals.
SDRi measures the difference between the SDR of the estimated target speech and that of the mixture.
An SDRi below 0 dB, indicates that extraction fails to emphasize the target signal, whereas an SDRi above 0 dB indicates that the target is emphasized compared to the mixture.
Values around 0 dB indicate that the mixture is extracted.
In contrast, large negative values would indicate that the interference signals are being extracted instead of the target.
All the experimental results were obtained by averaging the extraction performance of all the test utterances.

\subsection{Keyword-based method}
We compare the proposed ConceptBeam with a baseline method that extracts speech based on keywords obtained from object and speech recognition systems.
That is, this method extracts speech based on modality-specific recognition results instead of using modality-independent semantic information.

The left part of Figure~\ref{Baseline_and_separation_based_method}, shows the processing of the keyword-based method when an image is used as concept specifier.
First, we extract keywords from the concept specifier and the mixture.
We obtain the labels of the detected objects for the image specifier by performing object recognition on the image.
The labels serve as keywords.
We obtain the transcriptions from the speech signals (i.e., the mixture and the speech concept specifier) by performing automatic speech recognition (ASR).
We then remove the stop words from the transcriptions. The remaining words serve as keywords.
Note that we also keep time-alignment information between the speech signal and the transcriptions to know the starting and ending time of each keyword.
Then, we find the keywords associated with the mixture that match those in the concept specifier. Finally, we extract speech by applying a window on the segments around the detected keywords. Note that we set the window by adding a margin to the starting and ending time of each keyword to allow extracting the whole speech and not only the keywords.

For example, in Figure~\ref{Baseline_and_separation_based_method}, object recognition finds BOOK, MOUSE, and CAT in the image concept specifier. CAT and MOUSE match words in the speech recognition result, and speech is extracted by applying a window centered around these words.

\begin{figure}[!t]
  \centering
 \includegraphics[width=0.9\linewidth]{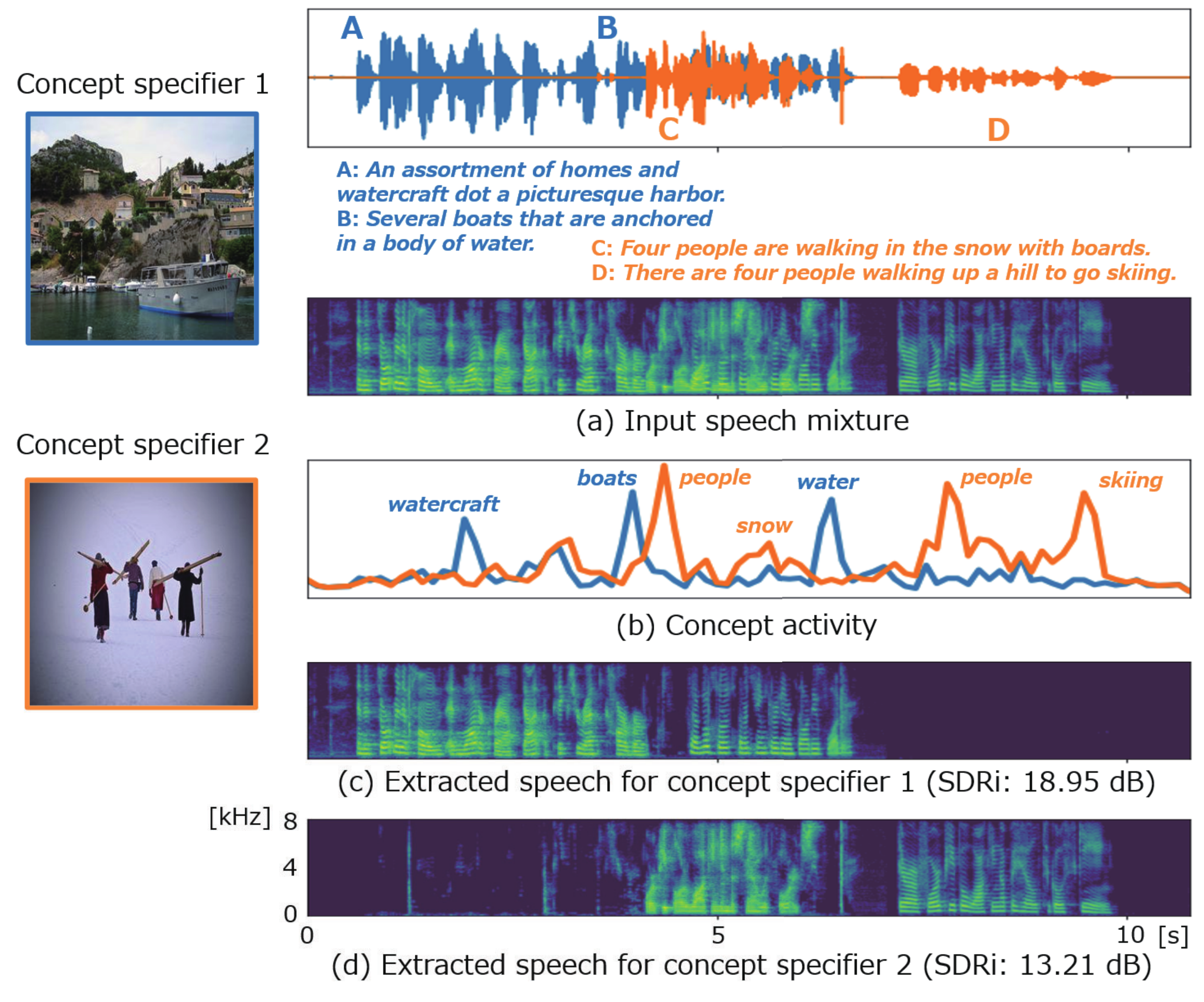}
  \vspace{-3mm}
  \caption{Example of speech extracted with ConceptBeam when concept specifiers are images.}
  \label{ConceptSpecifier_images}
  \vspace{-4mm}
\end{figure}

We used ESPnet \cite{watanabe2018espnet,arora2021espnet} for ASR, where we adopted a model\footnote{\url{Shinji Watanabe/librispeech_asr_train_asr_transformer_e18_raw_bpe_sp_valid.acc.best}} trained from the Librispeech corpus \cite{7178964}, which is publicly available on \texttt{ESPnet model zoo}.
The ``CTCSegmentation'' function was used to align the speech recognition result with the input speech signal.
We also used the Natural Language Toolkit (NLTK) to remove the stop words.
For object recognition, we used a Faster R-CNN model \cite{7485869} with a ResNet-50-FPN backbone implemented in Torchvision.
This model is pre-trained on COCO train2017, and the number of output classes is 91.
Based on a preliminary experiment, the margin used to define the windows around each keyword was set at 2 s.

\subsection{Separation-based method}
We also compare ConceptBeam with a separation-based method, which first separates the mixture into all source signals $\{\Vec{\tilde{x}}_1\ldots,\ \Vec{\tilde{x}}_L\}$ and then identifies the target speech among the separated signals using the concept specifier. 
The right side of Figure~\ref{Baseline_and_separation_based_method} shows such a separation-based method using an image as a concept specifier. 

\begin{figure}[t]
  \centering
 \includegraphics[width=0.9\linewidth]{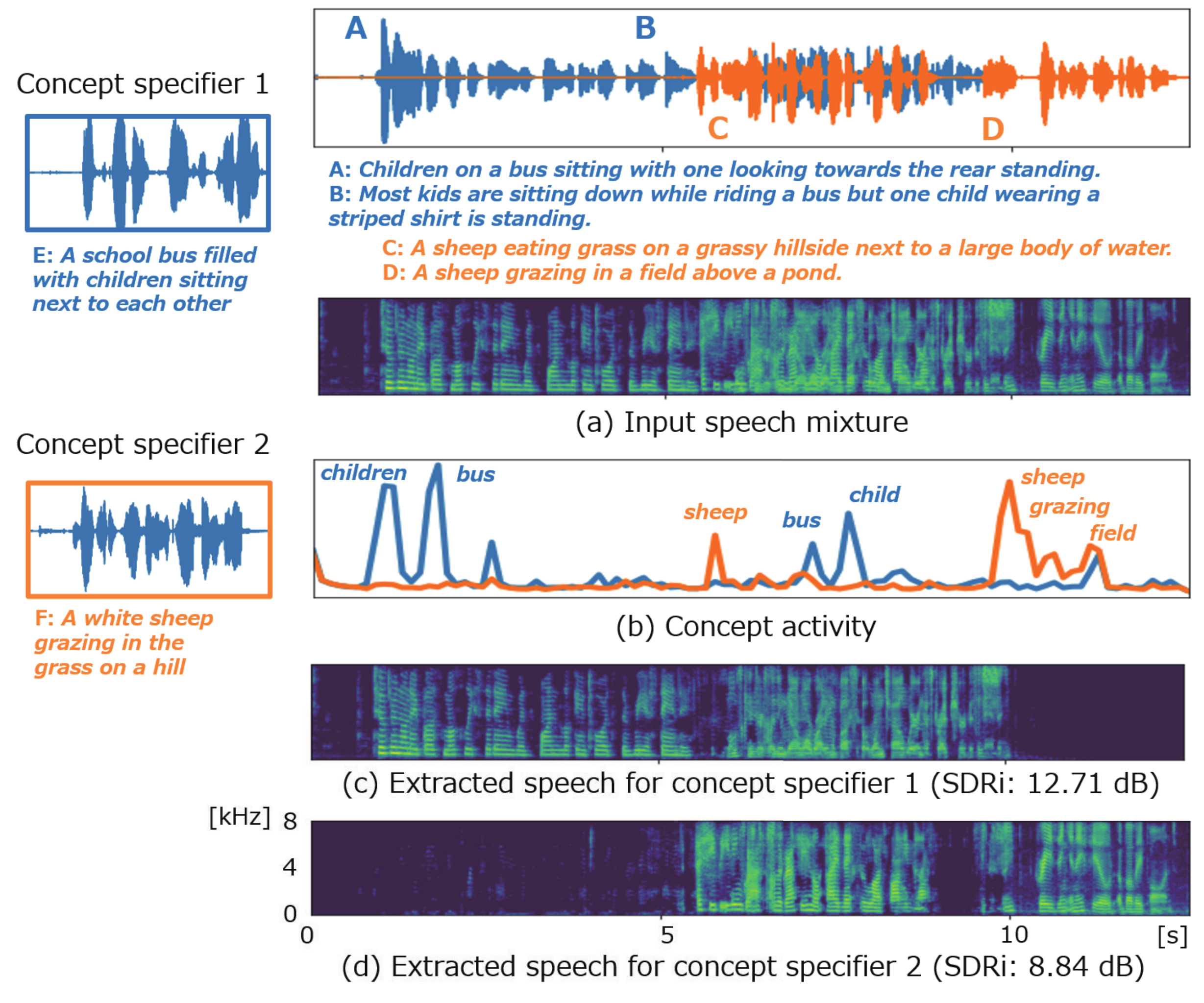}
  \vspace{-3mm}
  \caption{Example of speech extracted with ConceptBeam when concept specifiers are speech signals.}
  \label{ConceptSpecifier_speech}
  \vspace{-5mm}
\end{figure}

We use an NN-based separation method trained using permutation invariant training (PIT) \cite{Kolbak2017}. 
To identify the target concept, we first process each separated signal and the concept specifier $\Vec{\tilde{x}}_l$ and $\Vec{a}^c$ with the audio and concept encoders and obtain the semantic embeddings $\Vec{A}_l$ and $\Vec{E}^c$.
Figure~\ref{Baseline_and_separation_based_method} shows the case in which the concept specifier is an image.
We then determine the separated signals that contain the target concept based on the dot product of the embedding vectors, which is computed as
\begin{align}
    \Vec{\hat{x}}^c 
    &=\sum_l \alpha_l \tilde{x}_l,\quad \alpha_l=
    \begin{cases}
    1 & \text{if $\bar{\Vec{E}^c} \cdot \Vec{\bar{A}}_{l} > \theta$,} \\
    0 & \text{if $\bar{\Vec{E}^c} \cdot \Vec{\bar{A}}_{l} \leq\theta$,} 
  \end{cases}
  \label{eq:separation_based}
\end{align}
where $\bar{\Vec{E}^c}$ and $\bar{\Vec{A}}_{l}$ are obtained by averaging over the time or spatial dimension, $\Vec{E}^c$ and $\Vec{A}_{l}$, respectively.
The architecture of the source separation is the same as the extraction network of ConceptBeam except that the output layers consist of one output for each source in the mixture, where we assume the number of sources $L$ is equal to the number of utterances.
Therefore, this configuration requires knowing the number of speakers in the mixture.
Since this method is based on sound source (speaker) separation, Scenario 3 in Figure~\ref{fig:figure1} cannot be addressed.
In the training, the number of sources was set to $L\!=\!2$ for SpokenCOCO-2mix and $L\!=\!4$ for SpokenCOCO-4mix.
Based on the equal error rate obtained from the validation set, we set the hyperparameter, $\theta$, to 22.5 when the concept specifier is an image and to 8.9 when it is speech.

\begin{table*}
  \caption{SDRi for different settings of keyword-based and separation-based baseline, and the proposed ConceptBeam. 2mix and 4mix denote SpokenCOCO-2mix and SpokenCOCO-4mix, respectively. Bold font indicates the best performance. Percentage in parentheses indicates percentage of signals that contain speech, i.e., not only zeros. 
  }
\vspace{-3mm}
  \label{tab:comparison}
  \begin{tabular}{cccccrrrr}
    \toprule
    Method& \multicolumn{2}{c}{Testing set} &  \multicolumn{2}{c}{Concept specifier} & \multicolumn{4}{c}{Overlapping ratio} \\
     & {\it 2mix} & {\it 4mix}& {\it image} & {\it speech}  & \multicolumn{1}{c}{100\%} &  \multicolumn{1}{c}{50\%} & \multicolumn{1}{c}{25\%} & \multicolumn{1}{c}{0\% (non-overlap)}\\
    \midrule
    Keyword-based & $\checkmark$  & & $\checkmark$ &  & -1.43 (23.23\%) & 1.34 (39.92\%) & 5.65 (46.75\%) & 16.24 (48.74\%) \\
    Separation-based & $\checkmark$       & & $\checkmark$ & & \textbf{5.87} (82.21\%) & 8.60 (88.80\%) & 13.47 (91.51\%) & 16.80 (92.87\%)  \\
    ConceptBeam & $\checkmark$ & & $\checkmark$ & &  3.21 (100.0\%) & \textbf{10.33} (100.0\%) & \textbf{17.28} (100.0\%) & \textbf{19.15} (100.0\%) \\
    \midrule
    Keyword-based & $\checkmark$  & & & $\checkmark$ & -1.31 (48.25\%) & 1.14 (69.75\%) & 4.42 (76.82\%) & 13.61 (79.17\%)\\
    Separation-based & $\checkmark$       & &  & $\checkmark$ & 4.81 (76.19\%) & 7.85 (69.36\%) & 11.85 (61.14\%) & 14.37 (51.41\%) \\
    ConceptBeam & $\checkmark$ & & & $\checkmark$ &  \textbf{5.15} (100.0\%) & \textbf{11.39} (100.0\%) & \textbf{18.18} (100.0\%) & \textbf{19.90} (100.0\%) \\
    \midrule
    Keyword-based &  &$\checkmark$ & $\checkmark$ & & -4.51 (39.10\%) & -3.03 (51.90\%) & -1.52 (57.41\%) & 5.29 (61.08\%) \\
    Separation-based &   & $\checkmark$ & $\checkmark$ & & 0.56 (96.90\%) & 1.03 (98.65\%) & 2.97 (98.92\%) & \textbf{10.18} (99.37\%)  \\
    ConceptBeam &  & $\checkmark$ &$\checkmark$ & &  \textbf{4.59} (100.0\%) & \textbf{3.98} (100.0\%) & \textbf{3.40} (100.0\%) & 8.14 (100.0\%) \\
    \midrule
    Keyword-based &  &$\checkmark$ &  &$\checkmark$ & -3.46 (71.55\%) & -1.42 (83.11\%) & 0.34 (88.32\%) & \textbf{9.80} (91.18\%) \\
    Separation-based &   & $\checkmark$ &  & $\checkmark$& -2.12 (28.96\%) & -0.77 (19.15\%) & -1.65 (14.09\%) & -0.08 (11.71\%)  \\
    ConceptBeam &  & $\checkmark$ & &$\checkmark$ &  \textbf{5.42} (100.0\%) & \textbf{3.77} (100.0\%) & \textbf{3.45} (100.0\%) & 6.68 (100.0\%) \\
    \bottomrule
\end{tabular}
  \vspace{-2mm}
\end{table*}

\subsection{Results}


First, we illustrate the basic operation of ConceptBeam with the examples shown in Figures~\ref{ConceptSpecifier_images} and~\ref{ConceptSpecifier_speech}. 
In each figure, the waveform and spectrogram of an input speech mixture taken from SpokenCOCO-4mix are shown in (a).
Each colored waveform contains speech from two speakers (A and B or C and D) speaking about a particular concept.
As seen in the figures, the colored waveforms overlap.
We show for reference the text of each utterance, i.e., the original text captions.
The concept activity obtained with the image or speech concept specifiers by using Eq. \eqref{eq:concept_activity} are shown in (b).
We can see that the concept activity shows peaks for speech regions related to the concept.
These peaks can be related to the salient objects or words in the concept specifiers.
Note that we confirm with these examples that we can detect the concept activity even for the overlapping regions.
The spectrograms of the speech signals extracted from the input speech using the concept activities are shown in (c) and (d).
We can see that ConceptBeam successfully extracts speech related to the concept specifier with SDRi over 8.8 dB.

Table~\ref{tab:comparison} shows the extraction performance in terms of SDRi. 
With the keyword-based method, when the keywords obtained from the concept specifier do not match the keywords obtained from the input speech by ASR, the output signal is all zero. It is the same with the separation-based method when it is determined that all the separated signals do not contain any concepts obtained from the concept specifiers. This can be interpreted as missed detection errors. We report in parenthesis the percentage of signals that contain speech, i.e., not only zeros. 
Note that the reported SDRi results are computed using the output signals that are not zero signals since the SDR definition in \cite{Vincent2006} is ill-defined for zero signals.


We observe that the keyword-based method extracts the target speech relatively well when there is no overlap with SDRi of up to 16 dB. Performance drops dramatically when speech overlaps, because ASR performance degrades under overlapping conditions. Besides, the percentage of non-zero signals ranges between 23 and 91\%, which shows that it often cannot detect the concepts. The separation-based baseline achieves higher performance than the keyword-based baseline, especially under overlapping conditions, and detects the concepts better in most cases. 
However, since the separated signals are distorted as the number of speakers and the overlapping ratio increase and it is sensitive to the threshold $\theta$ of Eq. \eqref{eq:separation_based}, the performance drops significantly when using SpokenCOCO-4mix and the speech concept specifier.

In contrast, ConceptBeam clearly outperforms two baseline methods under most conditions. Note that ConceptBeam does not output zero signals because it does not perform hard-decision about the presence of the target concept.
Since speech generally tends to be sparse on a spectrogram, especially during voiced segments, even if the overlapping ratio is 100\%, the speech in the mixture does not always fully overlap.
As shown in Figures~\ref{ConceptSpecifier_images} and~\ref{ConceptSpecifier_speech}, the concept activity detection mechanism is robust to audio source overlap.
We consider a major reason for this to be that the concept generally corresponds to a longer signal period compared with a word or a signal frame for signal separation.
In other words, our concept-based method makes better use of such locally non-overlapping segments to separate the target signal.

\begin{figure}[t]
  \centering
 \includegraphics[width=0.95\linewidth]{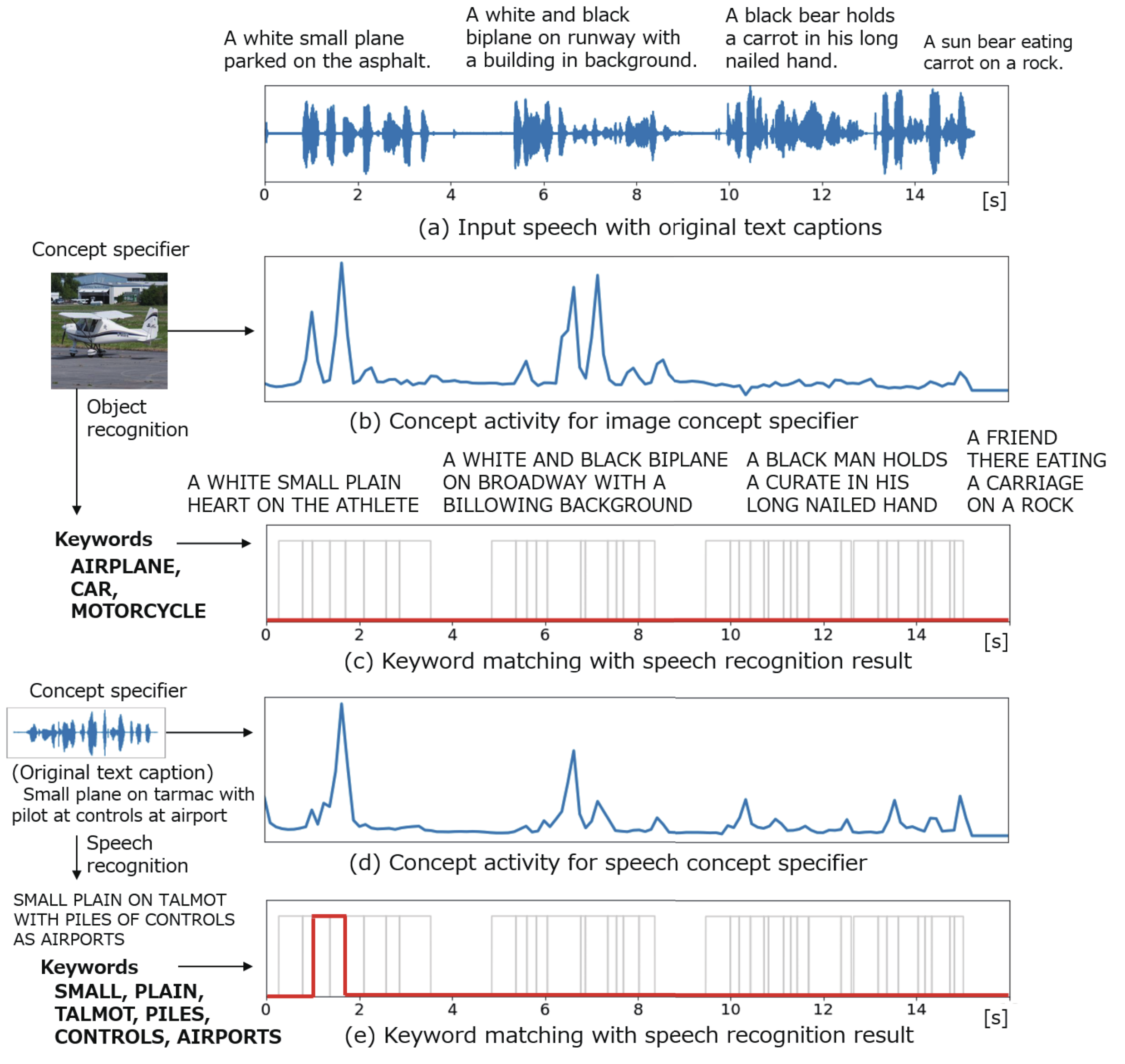}
  \vspace{-3mm}
  \caption{Concept activity of ConceptBeam and keyword matching of the keyword-based method.}
  \label{fig:Analysis}
  \vspace{-3mm}
\end{figure}

We take a closer look at the concept activity obtained with the keyword-based method and ConceptBeam to better understand the reasons for the missed detection errors of the keyword-based method.
Figure~\ref{fig:Analysis} shows an example of the concept activity obtained with the concept encoder and based on the keyword matching for a speech mixture without overlap.
It shows a typical case in which the keyword-based method fails to correctly identify the concept activity.
Figure~\ref{fig:Analysis}-(c) shows that the keywords obtained from the object recognition system do not match the text transcribed by ASR due to ASR errors or an expression variant (e.g., AIRPLANE vs BIPLANE).
Figure~\ref{fig:Analysis}-(e) shows that a part of the concept activity is detected, although the transcription is incorrect, because the ASR system makes the same error when recognizing the concept specifier and the mixture, i.e., it recognizes PLANE as PLAIN in both cases.

From Table~\ref{tab:comparison}, we find that the performance of ConceptBeam degrades particularly for SpokenCOCO-4mix compared to -2mix, even for the overlapping ratio of 0\%. We attribute this degradation to the fact that, as shown in Figure~\ref{dataset}, the target speech has a longer inactive region (or silence region) when there is less overlap.
The signals are also longer for SpokenCOCO-4mix than for -2mix, which implies that the inactive region is even longer. Since we trained ConceptBeam with only a highly overlapping mixture, i.e., overlapping ratio of 100\%, it was not trained to infer such long inactive regions. Furthermore, a concept activity sometimes involves false alarms. Consequently, ConceptBeam performs worse under these conditions.
In future work, we will investigate approaches to mitigate this issue \cite{kong2020source,borsdorf2021universal,wisdom2021what,Delcroix2022soundbeam}.

\begin{table}
  \caption{SDRi for mixtures consisting of all different speakers (Diff) and partly the same speakers (Same).}
  \label{tab:case3}
  \vspace{-2mm}
  \begin{tabular}{cccccrr}
    \toprule
     \multicolumn{2}{c}{Testing set} &  \multicolumn{2}{c}{Concept specifier} & \multicolumn{2}{c}{Non-overlap} \\
     {\it 2mix} & {\it 4mix}& {\it image} & {\it speech}  &  {\it Diff} & {\it Same}\\
    \midrule
     $\checkmark$  & & $\checkmark$ & &   \textbf{19.39} & 11.58 \\
      $\checkmark$       & &  & $\checkmark$ &  \textbf{20.20} & 10.62 \\
     &$\checkmark$ & $\checkmark$ & & \textbf{8.43}& 6.80\\
     &$\checkmark$ &  & $\checkmark$ & 6.59 &\textbf{9.43} \\
    \bottomrule
\end{tabular}
  \vspace{-2mm}
\end{table}


To confirm how well ConceptBeam can handle Scenario 3, in which one speaker speaks about two different concepts, in Figure~\ref{fig:figure1}, we took speech mixtures in which some of the spoken captions were uttered by the same speaker from the testing sets.
We obtained 589 out of 3,322 samples of SpokenCOCO-2mix and 104 out of 3,322 samples of Spoken COCO-4mix.
Table~\ref{tab:case3} compares the performance for the speech mixtures consisting of all different speakers (\textit{Diff}) and those partly containing the same speaker (\textit{Same}).
We observed positive SDRi for both conditions, which shows that ConceptBeam can extract speech based on the concept (not the speaker) even when the same speakers are speaking in the mixtures.
However, there is a large performance gap, which is not surprising as the extraction process relies mainly on the acoustic embedding derived from the concept activity.
We believe the acoustic embedding vector strongly contains speaker characteristics.
We plan to address this case by applying the concept activity to the output speech.

\begin{figure}[t]
  \centering
 \includegraphics[width=0.95\linewidth]{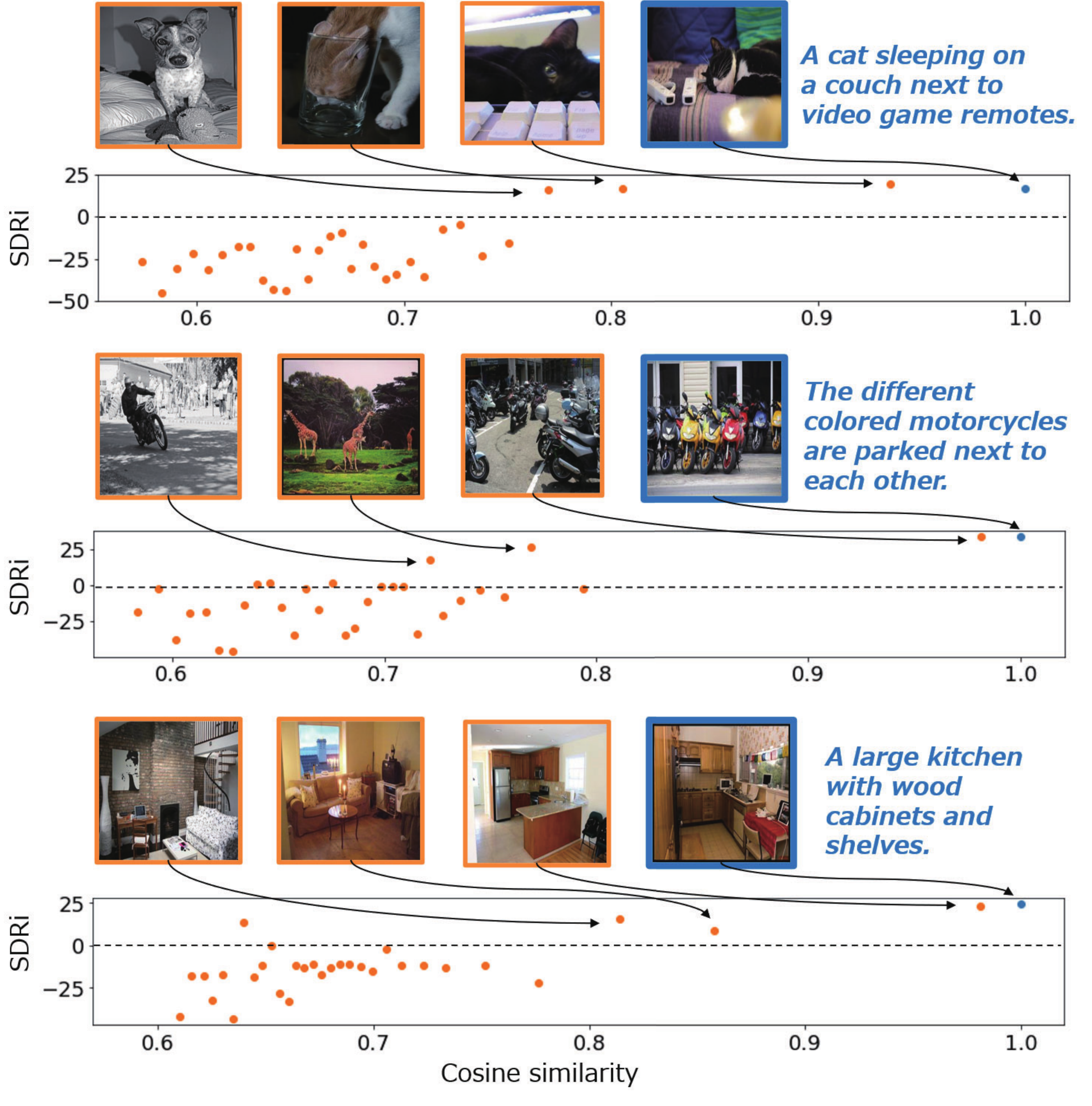}
  \vspace{-3mm}
  \caption{SDRi results for concept specifier images (orange dots) that are similar (in terms of cosine similarity in the semantic embedding space) to the image (blue dot) paired with the target speech to be extracted. Higher similarity, or a closer concept, tends to yield a higher SDRi.} 
  \label{fig:Similarity}
  \vspace{-2mm}
\end{figure}

Finally, we investigate the relationship between the SDRi and the similarity of the images to be used as the concept specifier. Here, the similarity is measured in terms of {\it concept} similarity.
If the proposed scheme effectively works, effective concept specifiers that are similar to each other will produce similar high SDRi scores, while others will not. 
To verify this, we selected image concept specifiers with concepts similar to the concept specifier paired with the target speech from the testing set, based on the cosine similarity between their semantic embeddings.
Figure~\ref{fig:Similarity} shows SDRi results of the extracted signals for the selected concept specifiers with the similarity scores (orange dots), where the blue dots show SDRi results for the original image concept specifiers.
It is shown that the target speech could be extracted for image concept specifiers with high similarity scores, while it is not extracted for concept specifiers with low similarity scores, as we expected. This demonstrates the effectiveness of the proposed scheme.

\section{Conclusion}
\label{sec:conclusion}
We proposed a novel framework, ConceptBeam, for target speech extraction based on semantic information.
We performed experiments with two types of concept specifiers, an image or speech, to describe concrete concepts such as physical objects or scenes.
The advantages of ConceptBeam are that it uses modality-independent semantic representation, making it independent of the word errors and expression variation issues of ASR-based systems and that it does not require the additional information on the number of speakers that the separation-based method does.
We anticipate that our proposed model will serve as a fundamental framework that could be used with other concept specifiers such as words and sentences to represent more abstract concepts.
The project page is available at \url{https://www.kecl.ntt.co.jp/people/ohishi.yasunori/project/conceptbeam/}.


\clearpage

\bibliographystyle{ACM-Reference-Format}
\balance
\bibliography{sample-base}

\end{document}